\documentclass[english]{article}
\usepackage{jcappub}
\usepackage{graphicx}

\begin{document}

\title{Machian gravity and a cosmology without dark matter and dark energy}
\author{Santanu Das}
\emailAdd{santanud@iucaa.ernet.in}
\affiliation{IUCAA, Post Bag 4, Ganeshkhind, Pune 411 007, India}

\abstract{
The standard model of cosmology is based on the general theory of relativity and demands more than 95\% of  the universe to consist of dark matter and dark energy that has no direct observational evidence till date. The foundation of the concept  these dark components are based on a fixed relation between the strength of the gravitational field and the matter density. 
Alternate models are put forward in past to explain the observations without dark components in the universe. Though they have their own merits and draw backs. In this paper we propose a new cosmological model based on Mach's principle. It provides a similar cosmology as that of the standard cosmological model without any ad-hoc dark matter or dark energy. We show that the theory naturally provides some geometric terms that behave like dark mater and dark energy and dark radiation.
The presence of dark radiation provides new observational features in cosmology. We show that the theory is supported by 
observational data from Big Bang Nucleosynthesis and Cosmic Microwave Background, and provides an explanations for excess number of effective neutrino species and higher Helium mass fraction in the universe. We also calculate the best fit cosmological parameters for our model using Planck+WP data. 
}

\maketitle

\section{Introduction}

General theory of Relativity proposed by Einstein relates the theory of gravity with the curvature of space-time. 
It described the orbits of planets up to an extraordinarily high precession and helped to provide the first satisfactory 
explanation to the perihelion shift of Mercury, which was earlier believed to be due to the presence of another unknown
planet or dust between Sun and Mercury \cite{Verrier2015}. Many of predictions of GR such as the bending of light and  
existence of gravitational wave etc. were later found to be true. All these contribute to the phenomenon success of the 
theory. However, when the theory is applied  in galactic and the extra galactic scale, 
it miserably fails to explain the observations, provided the calculations are done considering only the visible matter in the universe. This leads to
the concept of a new form of matter known as dark matter, consists of particles 
that interact with standard model particles only through
gravity.  In presence of dark matter the general theory of relativity can provide a proper description of gravity in galactic scale. 
Standard model of particle physics does not consists any candidate for dark matter. 
In extended particle physics models such as super-symmetric model, there are candidates for dark matter. However, high energy 
experiments don't provide any evidence for any such dark matter candidates till date. 

Recently WMAP-9 and Planck satellite measure the temperature fluctuations in the CMB sky up to few arc minute 
level of accuracy and the standard model of cosmology, based on general theory of relativity, explains the 
full observational data just by using a handful set of parameters. 
However, standard model of cosmology consider that almost 25\% of the universe consists of dark matter and 70\% of the 
rest consists of  dark energy. Dark energy is a hypothetical form of energy that provides negative pressure. 
Several models for dark energy are proposed in past, such as cosmological constant $\Lambda$, Quintessence~\cite{Zlatev1999,Steinhardt1999,Linder2005,Caldwell2002}, k-essence~\cite{Chimento2003,Jorge2007,dePutter2007,Das2013}, chaplain gas~\cite{Kamenshchik2001,Bento2002}, dark fluid~\cite{Arbey2005,Arbey2006,Arbey2008,Arbey2009} etc. that can explain the 
observational cosmology pretty well. However, the proposal of the dark matter or dark energy models are inductive, i.e. they are postulated to explain the galactic velocity curves or the accelerated expansion the the universe using the standard theory which otherwise fails to explain the observational data. The problem with the inductive science is that they always ignores the possibility of some fundamental flaw in the standard theory. Therefore, its important to check any such issue in the basic theory as some alternate theory may successfully explain 
the observational effects. 

Several theories are also proposed in past to provide an alternative explanation to the dark candidates in the universe. 
MOND~\cite{Milgrim1983,Milgrim1983a,Milgrim1983b} was proposed to explain the galactic rotation curves, but it violates momentum conservation. Theories like AQUAL~\cite{Bekenstein1984,Milgrom1986,Bekenstein2010} and TeVeS~\cite{Bekenstein2004,Bekenstein2009,Famaey2011,Mavromatos2009}  were proposed to provide MOND like field equations without violating any conservation laws. Another independent attempt was made by Moffat by proposing MoG/STVG~\cite{Moffat2006,Brownstein2005,Brownstein2005a,Brownstein2007,Moffat2007}
which can explain galactic velocity profiles, cluster mass profile, bullet cluster lensing effect and accelerated expansion of the 
universe without any dark components in the universe. All these theories modifies the Einstein-Hilbert Lagrangian with addition scalar and vector fields. On the other hand theories like $f(R)$ gravity~\cite{Buchdahl1970,Starobinsky1980,Sotiriou2008} are motivated from string theory, but requires extra dark matter to explain the observational data. Another brilliant theory based on Kaluza Klein theory in 5 dimension is  proposed by Wasson~\cite{Overduin1997,Wesson1992,Ponce1993} where matter is originated by the curvature of the space-time. Though the theory explains the cosmology very accurately, 
the origin of different kinds of particles are unclear in the theory.

An important issue that a gravity theory should address is the Mach's principle, a logical principle that relates the inertial properties of matter with the motion of distant objects of the universe. Though the general theory of relativity was founded based on Mach's principle, later it was realized that the GR does not follow the Mach's principle properly~\cite{Brans1962,Brans1977}. Attempts were made to formulate a new theory of gravity based on Mach principle, in Brans Dicke's varying $G$ model~\cite{Brans1961,Brans1962a}, Hoyle Narliker's Conformal gravity model~\cite{Hoyle1964} and by many others. Though none of these theories explain the cosmological observations without Dark components. Recently Moffat and Wasson shows that their STVG and Space-Time-Matter theories satisfy the Mach's principle~\cite{Moffat2007a,Mashhoon2011}.

We proposed new theory of gravity known as Machian Gravity in~\cite{Das2012,Das2012ab}, where it is shown that by introducing a fifth dimension we can explain the Mach's principle properly as well as it helps to link some of the quantum phenomenon with the time varying stochastic processes in the Universe. The field equations in the theory~\cite{Das2012ab} can explain the galactic velocity profiles, galaxy cluster mass, lensing phenomenon in the bullet cluster etc. properly without demanding any dark components in the universe. In this paper, we formulate cosmological model based on Machian Gravity and show that in four dimension it behaves identical to the standard cosmology model but without any ad-hoc dark matter or dark energy. Following the similar calculations as the Space-Time-Matter theory~\cite{Overduin1997,Wesson1992}, we  show that the dark matter and the dark energy in the theory are actually emerged from the geometry of the space-time and has nothing to do with the real matter. The theory also predict the existence of dark radiation. In this paper, we discuss some of the observational evidence for dark  radiation from Big Bang Nucleosynthesis and CMB data to support our theory .

The paper is organized as follows. The second section shows how a general five dimensional theory can be expanded in terms of
a four dimensional theory. In the third part, the most general line element for the cosmological model from the Machian gravity is taken and then it is shown how the dark matter and the dark energy can emerge from the geometry. In the fourth section we discuss the observational features of the theory that can distinguish our theory from the standard model of cosmology. The last section gives the conclusion. 

\section{General projection of 5 dimensional geometry into 4 dimension }

The 5 dimensional Minkowski line element used in Machian gravity~\cite{Das2012,Das2012ab} is given by $ds^2=c^2dt^2-dx^2-dy^2-dz^2-\epsilon\frac{\hbar^2}{4}d\zeta^2$, where $c$ is the speed of light and $\hbar=\frac{h}{2\pi}$ where $h$ is the Planck constant.
$\zeta$ is a new dimension which we termed as the background dimension and $\epsilon$ is its signature. Though logically the background 
dimension is consider to be space like~\cite{Das2012}, in this paper we explicitly show that a time-like ($\epsilon=-1$) extra dimension leads to results contradictory to the observations. We take $c=\frac{\hbar}{2}=8\pi G = 1$ for simplifying the calculations. 

A general five dimensional metric can be broken as a four-dimensional metric combined with a scalar field and a vector field as

\begin{equation}
\tilde{g}_{AB}=\left(\begin{array}{cc}
g_{\alpha\beta}+\kappa^{2}\phi^{2}A_{\alpha}A_{\beta} & \kappa\phi^{2}A_{\alpha}\\
\kappa\phi^{2}A_{\beta} & -\epsilon\phi^{2}
\end{array}\right)\,.
\end{equation}

\noindent Here $\phi$ is a scalar field, $A_{\alpha}$ is a four dimensional vector field, $\kappa$ is a constant and $g_{\alpha\beta}$ is the metric in the four dimension. We use tilde ($\tilde{\ldots}$) to denote the five dimensional tensors. The indices $A,B,\ldots$ are used to  represent a 5D entity and they vary from $0$ to $4$ and the indices $\alpha,\beta,\ldots$ vary from $0$ to $3$ and represent the 4D quantity.
 
Its difficult to work in a coordinate system where all the terms are present in the metric. Therefore, without loss of generality we choose a coordinate system where all the off-diagonal terms corresponds to the fifth dimension vanishes. This choice of coordinate system gives the line element as $ds^2 = g_{\alpha\beta}dx^{\alpha}dx^{\beta} - \epsilon\phi^{2}d\zeta^2$.

For this particular line element, we can calculate the 5D Reimann tensor and then express it in terms of a 4D Reimann tensor along with few additional terms. This gives us the following expressions~\cite{Wesson1992}

\begin{eqnarray}
\tilde{R}_{\alpha\beta} & = & R_{\alpha\beta}-\frac{\phi_{\alpha;\beta}}{\phi}-\frac{\epsilon}{2\phi^{2}}\Bigg(\frac{\partial_{4}\phi\partial_{4}g_{\alpha\beta}}{\phi}-\partial_{4}\partial_{4}g_{\alpha\beta}+g^{\mu\lambda}\partial_{4}g_{\alpha\mu}\partial_{4}g_{\beta\lambda}
 -\frac{1}{2}g^{\mu\nu}\partial_{4}g_{\mu\nu}\partial_{4}g_{\alpha\beta}\Bigg)\label{eq:R-alf-beta-a}\,, \\
\tilde{R}_{44} & = & \epsilon\phi g^{\mu\nu}\phi_{\mu;\nu}-\frac{1}{2}\partial_{4}g^{\mu\nu}\partial_{4}g_{\mu\nu}-\frac{1}{2}g^{\mu\nu}\partial_{4}\partial_{4}g_{\mu\nu}+\frac{1}{2\phi}\partial_{4}\phi g^{\mu\nu}\partial_{4}g_{\mu\nu}
 -\frac{1}{4}g^{\mu\nu}g^{\lambda\sigma}\partial_{4}g_{\lambda\mu}\partial_{4}g_{\sigma\nu}\label{eq:R-4-4}\,, \\
\tilde{R}_{4\alpha} &=& \phi \label{eq:R-4-alf} \left[\frac{1}{2\phi}\left(g^{\beta\lambda}\partial_{4}g_{\lambda\alpha}-\delta_{\alpha}^{\beta}g^{\mu\nu}\partial_{4}g_{\mu\nu}\right)\right]_{;\beta} \,,
\end{eqnarray}

\noindent where, $\tilde{R}_{\alpha\beta}$ is the first $4\times4$ components of the five dimensional Reimann tensor and 
$R_{\alpha\beta}$ is the four dimensional Riemann tensor calculated from $g_{\alpha\beta}$. 
In the above equations `$;$' represents the covariant derivative, 
 $\phi_{\alpha}=\partial_{\alpha}\phi$
and $\phi_{\alpha;\beta}=\left(\phi_{\alpha,\beta}-\phi_{\lambda}\Gamma_{\alpha\beta}^{\lambda}\right)$.

According to the field equation of Machian Gravity $\tilde{R}_{AB}=\tilde{T}_{AB}-\tilde{g}_{AB}\tilde{T}$~\cite{Das2012ab}, where $\tilde{T}_{AB}$ is the five dimensional stress energy tensor. In the vacuum it gives $\tilde{R}_{\alpha\beta}=0$. But it does not imply $R_{\alpha\beta}=0$ due to the presence of the additional terms given in Eq.(\ref{eq:R-alf-beta-a}). These extra terms come due to the nature of the reference frame. In an inertial (non accelerating) reference frame these terms are 0. However, in a non-inertial (accelerating) reference frame these terms will be nonzero. Therefore, we can get similar effect if a particle accelerate in the universe or if a particle is fixed and the entire universe starts accelerating towards it. This satisfies the basic 
concept of the Mach's principle.

The stress energy tensor in Machian Gravity is given by  $\tilde{T}_{AB}=(\rho+p)\tilde{u}_{A}\tilde{u}_{B}+p\tilde{g}_{AB}$~\cite{Das2012ab}. In the classical limit up to a very reasonable assumption we can assume $\tilde{T}_{\alpha\beta}\approx T_{\alpha\beta}$, because in the classical limit $\hbar\rightarrow0$. Also if $\tilde{T}_{A4}$ 
are not negligible in comparison to $\tilde{T}_{\alpha\beta}$ then the classical energy momentum conservation laws do not get satisfied. 
In the field equation we relate $\tilde{G}_{AB}=\tilde{R}_{AB}-\frac{1}{2}\tilde{g}_{AB}\,\tilde{R}$ with $\tilde{T}_{AB}$.
Using Eq.(\ref{eq:R-alf-beta-a}), Eq.(\ref{eq:R-4-alf}) and  Eq.(\ref{eq:R-4-4}) we can calculate $\tilde{G}_{AB}$ and express its 
first $4\times 4$ component as $\tilde{G}_{\alpha\beta} = G_{\alpha\beta} + Q_{\alpha\beta}$, where $G_{\alpha\beta}$ is the four 
dimensional Einsteins tensor and $ Q_{\alpha\beta}$ consists of the remaining terms that are not included in the four dimensional 
Einstein's tensor. Therefore the field equation $\tilde{G}_{AB}=\tilde{T}_{AB}$ leads to
\begin{equation}
G_{\alpha\beta}=T_{\alpha\beta}-Q_{\alpha\beta} \,. 
\end{equation}

\noindent If it can be shown that the terms $-Q_{\alpha\beta}$ for a five dimensional FLRW metric have the same property as that of 
dark components of the universe then the theory can predict everything in the same way as that of the standard 
cosmology without demanding any form ad.hoc. matter and energy components.

\section{Cosmological solution from a generalized metric}

A general spherically symmetric  line element can be taken as

\begin{equation}
ds^{2}=e^{\omega}dt^{2}-e^{\kappa}dr^{2}-R^{2}\left(d\theta^{2}+\sin^{2}\theta d\phi^{2}\right)-\epsilon e^{\mu}d\zeta^{2}\,.\label{eq:line-element-2}
\end{equation}

\noindent The exponentials are taken to ensure that these quantities can't be negative. 

For this particular line element  we can calculate the $\tilde{R}_{AB}$ using  Eq.(\ref{eq:R-alf-beta-a}), Eq.(\ref{eq:R-4-alf}) and  Eq.(\ref{eq:R-4-4})
and then $\tilde{G}_{AB}$ and express them in terms of $G_{\alpha\beta}$ along with some additional terms. This gives the following expressions~\cite{Overduin1997}

\begin{eqnarray}
\tilde{G}_{0}^{0} & = & G_{0}^{0}+e^{-\omega}\Bigg(\frac{\dot{\mu}\dot{\kappa}}{4}+\frac{\dot{\mu}\dot{R}}{R}\Bigg)-e^{-\kappa}\Bigg(\frac{R'\mu'}{R}-\frac{\kappa'\mu'}{4}+\frac{\mu''}{2}+\frac{\mu'^{2}}{2}\Bigg)\nonumber 
 -\epsilon e^{-\mu}\left(\frac{\kappa^{**}}{2}+\frac{\kappa^{*2}}{4}\right. \\
& &\left. -\frac{\kappa^{*}\mu^{*}}{4}+\frac{R^{*}}{R}\left(\kappa^{*}-\mu^{*}\right)+\frac{R^{*2}}{R^{2}}+\frac{2R^{**}}{R}\right)\label{eq:G-0-0}\,, \\
\tilde{G}_{0}^{1} &=& G_{0}^{1}+e^{-\kappa}\left(\frac{\dot{\mu}'}{2}+\frac{\dot{\mu}\mu'}{4}-\frac{\omega'\dot{\mu}}{4}-\frac{\dot{\kappa}\mu'}{4}\right)\label{eq:G-1-0}\,, \\
\tilde{G}_{1}^{1} & = & G_{1}^{1}+e^{-\omega}\left(\frac{\ddot{\mu}}{2}+\frac{\dot{\mu}^{2}}{4}-\frac{\dot{\omega}\dot{\mu}}{4}+\frac{\dot{R\dot{\mu}}}{R}\right)-e^{-\kappa}\left(\frac{\mu'\omega'}{4}+\frac{\mu'R'}{R}\right)\nonumber 
-\epsilon e^{-\mu}\left(\frac{\omega^{**}}{2}+\frac{\omega^{*2}}{4}\right. \\
& &\left.+\frac{R^{*2}}{R^{2}}+\frac{2R^{**}}{R}+\frac{R^{*}}{2R}\left(\omega^{*}-\mu^{*}\right)-\frac{\mu^{*}\omega^{*}}{4}\right)\,,\label{eq:G-1-1}
\end{eqnarray}

\begin{eqnarray}
\tilde{G}_{2}^{2} & = & G_{2}^{2}+e^{-\omega}\left(\frac{\dot{R}\dot{\mu}}{2R}-\frac{\dot{\omega}\dot{\mu}}{4}+\frac{\dot{\mu}\dot{\kappa}}{4}+\frac{\ddot{\mu}}{2}+\frac{\dot{\mu}^{2}}{4}\right)-e^{-\kappa}\Bigg(\frac{R'\mu'}{2R}+\frac{\mu''}{2}\nonumber 
 +\frac{\mu'^{2}}{4}+\frac{\omega'\mu'}{4} \\
 &  &-\frac{\mu'\kappa'}{4}\Bigg)
-\epsilon e^{-\mu}\Bigg(\frac{R^{**}}{R}+\frac{R^{*}\omega^{*}}{2R}+\frac{R^{*}\kappa^{*}}{2R}-\frac{R^{*}\mu^{*}}{2R}\nonumber 
 +\frac{\omega^{**}}{2}+\frac{\omega^{*2}}{4}+\frac{\kappa^{**}}{2}+\frac{\kappa^{*2}}{4} \\
 &  &+\frac{\kappa^{*}\omega^{*}}{4}-\frac{\kappa^{*}\mu^{*}}{4} 
-\frac{\mu^{*}\omega^{*}}{4}\Bigg)\label{eq:G-2-2}\,, \\
\tilde{G}_{3}^{3} &=& \tilde{G}_{2}^{2}\label{eq:G-3-3}\,.
\end{eqnarray}

\noindent Here $\dot{(\ldots)}$, $(\ldots)'$  and  $(\ldots)^{*}$ represent the derivative with respect
to the $t$, $r$ and $\zeta$ respectively.

If the extra geometric terms in the right hand side of the expressions are treated as 
the stress energy tensor for a fluid in equilibrium
then it will have $\tilde{u}_{0}\neq0$ and $\tilde{u}_{1}\neq0$ and $\tilde{u}_{2}=\tilde{u}_{3}=0$.
If $\tilde{u}_{2}$ or $\tilde{u}_{3}$ were nonzero then we should
have some nonzero $\tilde{G}_{2}^{0}$ or $\tilde{G}_{2}^{1}$, 
which is not the case. Also we consider $\tilde{u}_{4}$ to be negligible in the classical limit. 

The negative values of the expressions in the right hand side of Eq.(\ref{eq:G-0-0})
-Eq.(\ref{eq:G-3-3}) will have the terms having the properties of Dark matter and Dark energy. 
We can consider these expressions as the expressions for $\tilde{Q}^{\mu}_{\nu}$. These $\tilde{Q}^{\mu}_{\nu}$ should give the stress energy tensor for the dark components of the universe. These are purely geometric terms and has nothing to do with the real matter and energy. If we define the density and the pressure from these geometric
components by $\rho_{g}$ and $p_{g}$ respectively, then these two components can be calculated from the stress energy tensor as 

\begin{equation}
\rho_{g}=\tilde{Q}_{0}^{0}+\tilde{Q}_{1}^{1}-\tilde{Q}_{2}^{2}\label{eq:delsity}\,,
\end{equation}

\noindent and 

\begin{equation}
p_{g}=-\tilde{Q}_{2}^{2}\label{eq:pg}\,.
\end{equation}

\noindent Here we use $\tilde{u}^{1}\tilde{u}_{1}+\tilde{u}^{2}\tilde{u}_{2}=1$, as other components
of velocity are zero. The subscript $g$ represents that these terms are purely geometric. 

As $\tilde{T}_{44}\sim 0$ in the classical limit., we can also consider $R_{44}\sim0$ and so is $R_{4}^{4}$. Hence adding $R_{4}^{4}$ with any other expression will not change the value of that expression. The expression for $R_{4}^{4}$ is given by

\begin{eqnarray}
\tilde{R}_{4}^{4} & = & -\epsilon e^{-\mu}\left(-\frac{\omega^{**}}{2}-\frac{\omega^{*2}}{4}-\frac{\kappa^{**}}{2}-\frac{\kappa^{*2}}{4}+\frac{\mu^{*}\omega^{*}}{4}+\frac{\mu^{*}\kappa^{*}}{4}+\frac{\mu^{*}R^{*}}{R}
-\frac{2R^{**}}{R}\right)- e^{-\omega}\left(\frac{\ddot{\mu}}{2}+\frac{\dot{\mu}^{2}}{4}\right. \nonumber \\ 
& & \left.-\frac{\dot{\mu}\dot{\omega}}{4}+\frac{\dot{\mu}\dot{\kappa}}{4}+\frac{\dot{\mu}\dot{R}}{R}\right)
 + e^{-\kappa}\left(\frac{\mu''}{2}+\frac{\mu'^{2}}{4}+\frac{\mu'\omega'}{4}-\frac{\mu'\kappa'}{4}+\frac{\mu'R'}{R}\right)
\end{eqnarray}

A little algebraic manipulation can show that if $R_{4}^{4}$ is added to Eq.(\ref{eq:pg}) then the expression for $p_{g}$  becomes highly simplified and more meaningful. Thus, expression for $p_{g}$ is taken as

\begin{equation}
p_{g}=-\tilde{Q}_{2}^{2}+\tilde{R}_{4}^{4}\label{eq:pressure}\,.
\end{equation}

Now putting the expressions for $\tilde{Q}_{0}^{0}$, $\tilde{Q}_{1}^{1}$ and $\tilde{Q}_{2}^{2}$ from Eq.(\ref{eq:G-0-0}), Eq.(\ref{eq:G-1-1}) and Eq.(\ref{eq:G-2-2}) into Eq.(\ref{eq:delsity}) and Eq.(\ref{eq:pressure}), the expressions for the density and pressure can be calculated as 

\begin{eqnarray}
\rho_{g} & = & \frac{3}{2}\left(\frac{e^{-\kappa}\mu'R'}{R}-\frac{e^{-\omega}\dot{\mu}\dot{R}}{R}\right)-\frac{3}{2}\epsilon e^{-\mu}\left(\frac{R^{*}\mu^{*}}{R}-\frac{2R^{**}}{R}\right)+\epsilon e^{-\mu}\frac{R^{*2}}{R^{2}}-\epsilon e^{-\mu}\left(\frac{\omega^{*}\kappa^{*}}{4}\right)\nonumber \\
 &  &+\epsilon e^{-\mu}\frac{R^{*}}{2R}\left(\kappa^{*}+\omega^{*}\right)\label{eq:density_1}\,, \\
p_{g} & = & \frac{1}{2}\left(\frac{e^{-\kappa}\mu'R'}{R}-\frac{e^{-\omega}\dot{\mu}\dot{R}}{R}\right)-\frac{1}{2}\epsilon e^{-\mu}\left(\frac{R^{*}\mu^{*}}{R}-\frac{2R^{**}}{R}\right)-\epsilon e^{-\mu}\left(\frac{\omega^{*}\kappa^{*}}{4}\right)\nonumber \\
& &-\epsilon e^{-\mu}\frac{R^{*}}{2R}\left(\kappa^{*}+\omega^{*}\right)\label{eq:pressure_1}\,.
\end{eqnarray}

\noindent These expressions show that there are four different types of components of the pressure and density in the above equations. These can be broken down as follows 

\begin{eqnarray}
\rho_{gr}&=&3p_{gr}=\frac{3}{2}\left(\frac{e^{-\kappa}\mu'R'}{R}-\frac{e^{-\omega}\dot{\mu}\dot{R}}{R}\right)-\frac{3}{2}\epsilon e^{-\mu}\left(\frac{R^{*}\mu^{*}}{R}-\frac{2R^{**}}{R}\right)\label{eq:radiation}\,, \\
\rho_{gd}&=&\epsilon e^{-\mu}\frac{R^{*2}}{R^{2}}\label{eq:dust}\,, \\
\rho_{gs}&=&p_{gs}=-\epsilon e^{-\mu}\left(\frac{\omega^{*}\kappa^{*}}{4}\right)\label{eq:stiff-matter}\,, \\
\rho_{g\Lambda}&=&-p_{g\Lambda}=\epsilon e^{-\mu}\frac{R^{*}}{2R}\left(\kappa^{*}+\omega^{*}\right)\label{eq:dark-energy}\,. 
\end{eqnarray}

\noindent
The pressure and density of the quantity given by Eq.(\ref{eq:radiation}), follows the relation $p=\frac{\rho}{3}$ and they behaves exactly as photon or massless neutrino in the universe, provided they interact with other particles through the space-time curvature or gravity. Hence, this component can be treated as a dark radiation in the standard cosmology. The presence of the dark radiation has been postulated literature by many authors, the detail of which is discussed in the next section. 
The $2^{nd}$ component, given by Eq.(\ref{eq:dust}) behaves as a non-relativistic matter, with $0$ pressure. This fulfills all the
criteria for the cold dark matter. The third component, i.e. Eq.(\ref{eq:stiff-matter})
is another interesting component, where pressure and the density are equal. This is the stiffest equation of state that a fluid can have, because after this the speed of sound inside a fluid will exceed the speed of light, violating the consistency relation. This kind of fluid was once proposed by Zeldovich and named as stiff matter. The last component, given by Eq.(\ref{eq:dark-energy}) has the $\rho=-p$. Thus it behaves as dark energy of the standard cosmological model. 

In Machian gravity dark matter and dark energy emerges automatically from the geometry. Thus, the theory can provide a cosmological model exactly similar to that of the standard cosmological model without demanding any external dark mater or
dark energy. Also Eq.(\ref{eq:radiation})- Eq.(\ref{eq:dark-energy}) shows that $\epsilon=-1$ gives a negative negative energy density for dark matter which is completely illogical. It provides a completely independent prove that  the extra dimension should be space-like.

For cosmology we need to use the FLRW line element for the 1+3 dimension. Thus the line element in the five dimensional universe should be written as

\begin{equation}
ds^{2}=dt^{2}-a^{2}(t,\zeta)\left[dr^{2}+r^{2}\left(d\theta^{2}+\sin^{2}\theta d\phi^{2}\right)\right]-e^{\mu}d\zeta^{2}\label{eq:line-element-3}\,.
\end{equation}

\noindent Comparing this line element with Eq.(\ref{eq:line-element-2}) we get $e^{\omega}=1$  and hence $\omega^{*}$  vanishes. This gives the stiff matter energy density to be zero. Therefore, no stiff matter component will be present in standard model of cosmology. 

\section{Differentiating the model from standard cosmology using observation}

The standard model of cosmology requires cold dark matter and dark energy to explain the observational data set. It does not require any component of dark radiation. However, in the model proposed in this paper, a new component of dark radiation emerge out from the theory that interacts with other matter components only through gravity. In the standard cosmological model the massless neutrinos are very weakly interacting particles and after decoupling at the very early universe they also interact with other particles only through gravity. No other interactions are significant for them. Therefore, the presence of the dark radiation in the universe can be modeled as the presence of the excess Neutrinos in the universe. 

Recently the results from WMAP, Planck and other experiments shows that in the best fit $\Lambda$CDM model the effective number of neutrino species is   $\sim4.0$~\cite{Das201a}, which is more than its theoratical value of $3.047$.  
This presence of excess number of neutrino species in the universe leads researchers to look for the existence of dark radiation in the recent literature. Ackerman et.al. proposed a new long range $U(1)$ gauge field that coupled to the dark matter but not to the standard model particles and explore its consequences~\cite{Ackerman2009}. Many other theoretical models for dark radiation~\cite{Archidiacono2013,Harko2006,Hasenkamp2013} are also recently explored in the the literatures. However, in the cosmology model  proposed in this paper,  we get the dark radiation as a integral part of the theory. This distinguishes our model from standard cosmology. 

The existence of the dark radiation in the universe affects the Big Bang Neucliosynthesis, CMB, the structure formation and many other phenomenon of the universe.

\subsection{Big Bang Neucliosysthesis}

During the Nucleosynthesis production of proton were favored over neutron due to their lesser mass. Free neutrons decay to protons through $\beta$ decay with a half life of around $10.2$ min. Presence of dark radiation increases the expansion rate of the universe in the radiation dominated era giving lesser time for the Nucleosynthesis and less time to convert neutrons to protons. This will provide a increased neutron to proton ratio in the universe. We can see its effects by measuring the 
primordial Helium fraction. In presence of the dark radiation the primordial Helium fraction ($Y_{\rm He}$) increases from its theoretical 
value of $0.24$. 

The constraints on the Helium mass fraction can be directly derived from the low metalicity extragalactic H-II regions. A recent  analysis of a sample of 93 spectra of 86 low-metalicity extragalactic regions put a constrain of primordial helium fraction $Y_{\rm He}=0.2565\pm0.006$ and the effective number of neutrinos corresponding to it is $N_{\nu}=3.80^{+0.80}_{-0.70}$~\cite{Izotov2010,Steigman2012}. This excess number of ($\Delta N_{\nu} = 0.76$) neutrino species can be treated as a quantifier for the dark radiation. A 7 parameter analysis of CMB power spectrum using Planck+WP data also constrain $Y_{\rm He}=0.256\pm0.016$~\cite{SCoPE}. These hints towards the existence of dark radiation in the universe. 
\begin{figure}
\centering
\includegraphics[width=0.6\textwidth,trim = 60 260 70 280, clip]{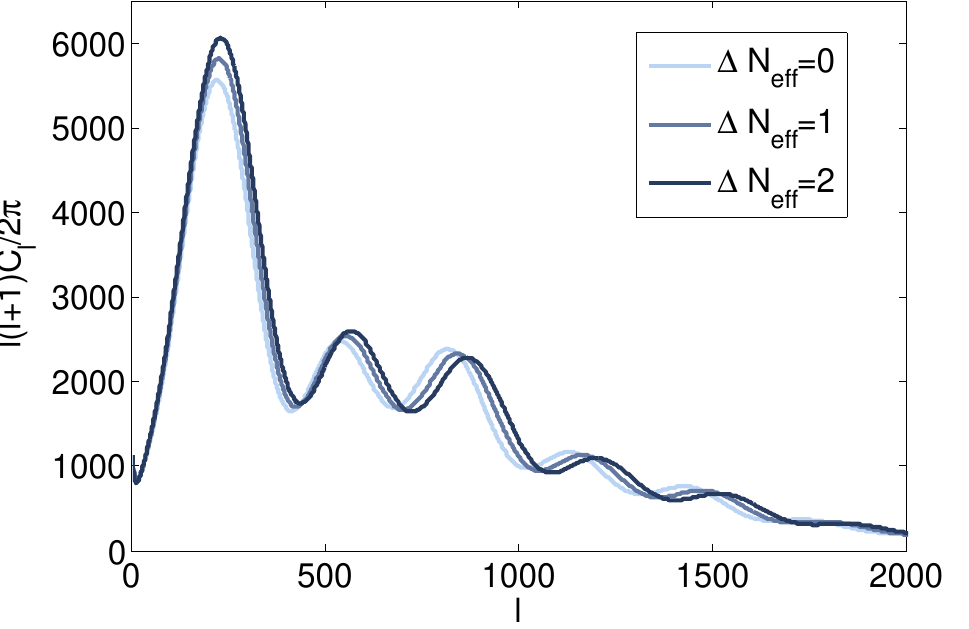}          
\caption{\label{fig:Cl_N_eff}Effect of increasing the number of Neutrino species ($N_{\nu}=3.04+\Delta N_{\nu}$) in cosmological power spectrum. With the increase in the effective number of Neutrinos the power in the first two peaks of the CMB power spectrum increases due to early ISW effect  and the power in rest of the peaks decreases due to enhanced Silk damping.}
\end{figure}

\subsection{Effects on CMB}

The effect of dark radiation in CMB is two fold. The ISW effect of the CMB power spectrum is directly linked with the amount of matter. So if we increase the effective number of neutrinos or if we add extra dark radiation component then the power in the ISW effect, specially the early ISW effect will increase. Therefore, the power in the first two peaks in the CMB power spectrum will increase. However, increase of the dark radiation component will also increase the silk damping. Therefore, the higher multipoles of the CMB power spectrum will be damped~\cite{Archidiacono2013}. In figure(\ref{fig:Cl_N_eff}) this particular effect is shown using excess of number of neutrino species as a quantifier for dark radiation in the universe.

The number of the effective neutrino species form Planck, WP and  BICEP-2 experiments are discussed in~\cite{Das201a}. With Planck+WP likelihood the effective number of neutrino is $N_{\nu}=3.4867\pm0.31$. and with Planck+WP+lensing likelihood $N_{\nu}=3.447\pm0.34$. If we add BICEP likelihood then the the effective number of neutrinos tends to $4.0$. The increase of the  effecting number of neutrinos from the expected standard model neutrino number shows the presence of the dark radiation in the universe. 

\begin{figure}
\centering
\includegraphics[width=0.85\textwidth,trim = 0 200 5 180, clip]{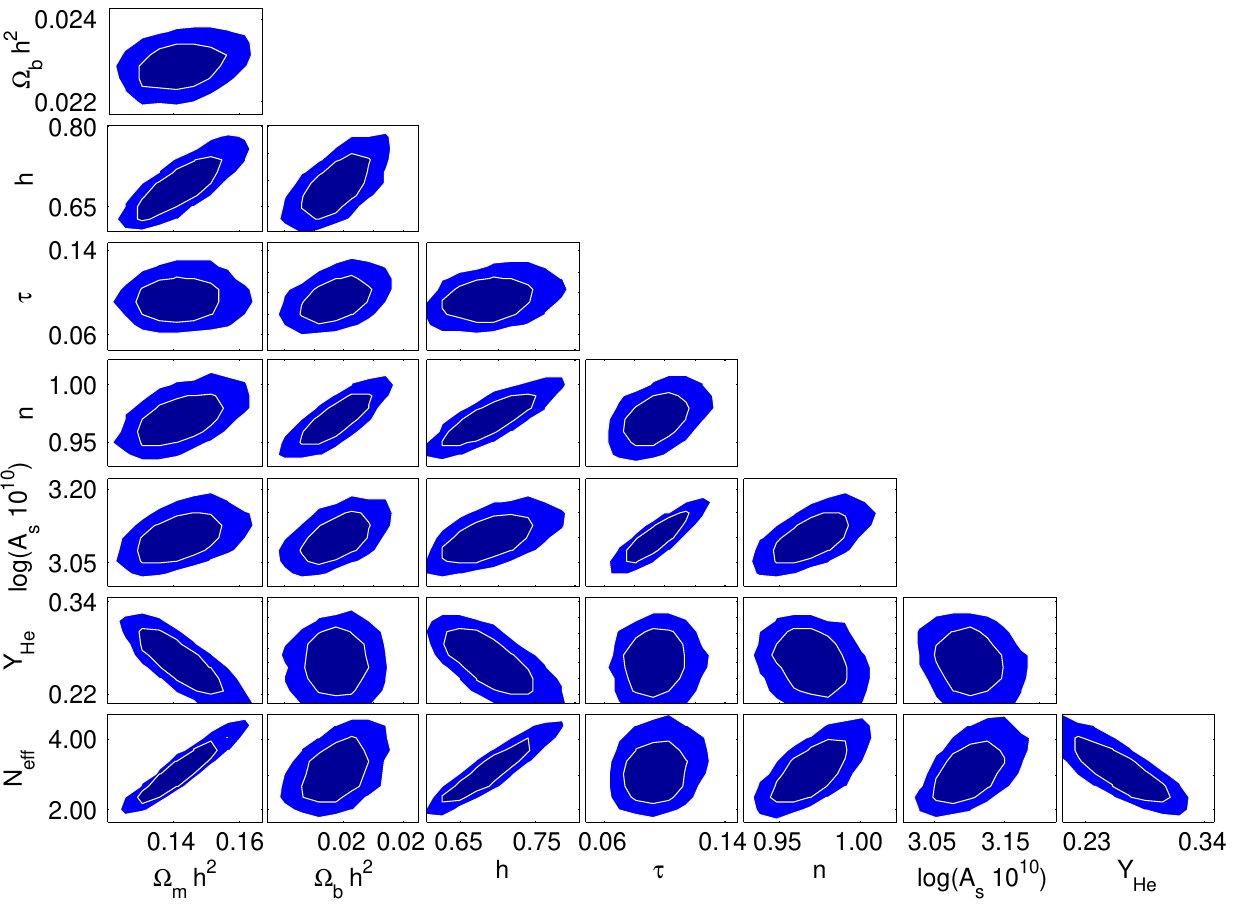}          
\includegraphics[width=0.85\textwidth,trim = 10 250 5 250, clip]{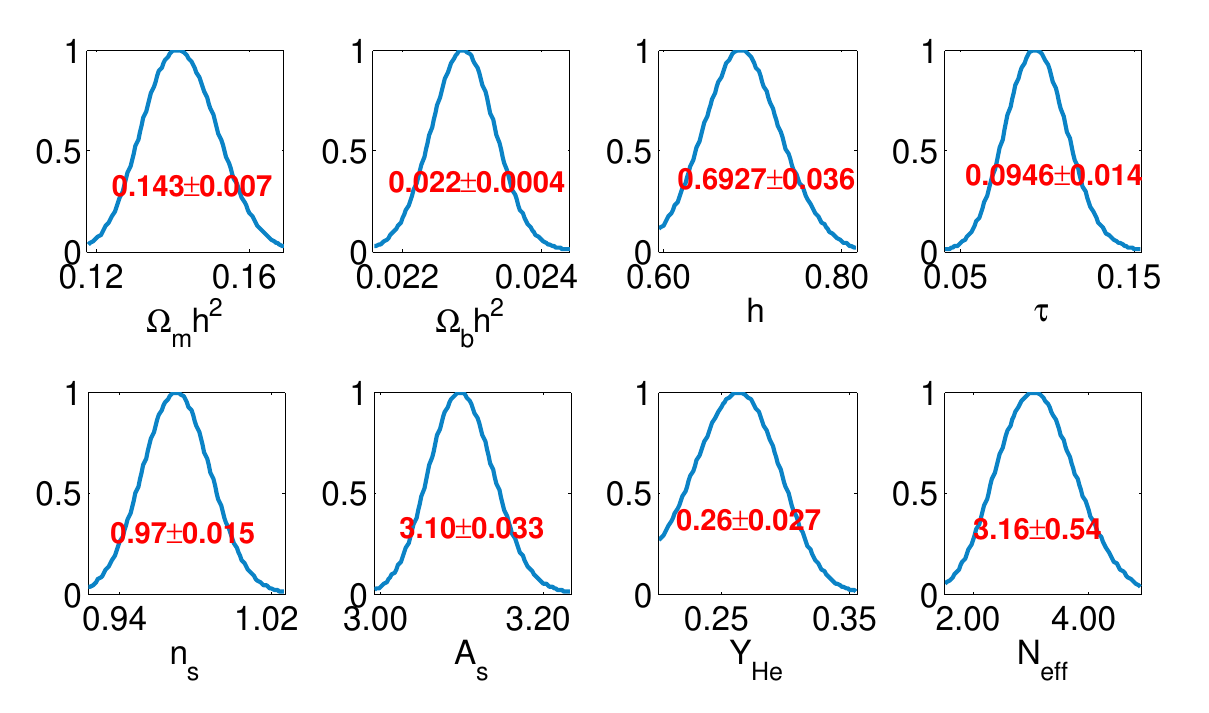}          
\caption{\label{fig:two}Top: Two dimensional likelihood contours  for Planck+ WP likelihood with 8 parameter ($\Omega_m h^2$, $\Omega_b h^2$, $h$, $\tau$, $n_s$, $A_s$, $Y_{\rm He}$ and $N_{\nu}$) model. Bottom: The one dimensional marginalized probability distribution plots for the parameters. The average value and standard deviation of the parameters are shown in red. The plots show a strong negative correlation between $Y_{\rm He}$ and $N_{\nu}$. Average value of $N_{\nu}$ is larger than the effective theoretical value of $3.047$. and $Y_{\rm He}$ is higher than its theoretical value of $0.24$.}
\end{figure} 

\subsection{Estimating the cosmological parameters}

Here we provide an estimate of the cosmological parameters for a 8 parameter model from Planck and WMAP polarization likelihood where we consider $N_{\nu}$ and $Y_{\rm He}$ as parameters along with the 6 standard parameters i.e. $\Omega_m h^2$, $\Omega_b h^2$, $h$, $\tau$, $n_s$, $A_s$. The Helium mass fraction $Y_{\rm He}$ is dependent on the effective number of neutrinos. Therefore, they are not completely independent parameters. Also increasing $Y_{\rm He}$ has an opposite effect on the high multipoles of the CMB power spectrum than that of the $N_{\nu}$. Therefore, if we constrain $N_{\nu}$ and $Y_{\rm He}$ from CMB power spectrum then they will counter each others effect and come closer to the standard model value. However, such an analysis can provide us the constrain on these parameters from CMB alone. In figure(\ref{fig:two}) we show the constraints on these parameters from the 8 parameter analysis. Our analysis show that even though the $Y_{\rm He}$ and $N_{\nu}$ nullify each other's effect, their average values are higher than the theoretical values hinting towards the presence of dark radiation in the universe.

\section{Conclusion}

We present a new cosmology model for Machian gravity theory and show that the dark matter and the dark energy in this theory emerges from the space-time geometry. This eliminates the requirement for any additional matter or energy component in the universe for explaining the observational dataset. Our cosmological model behaves exactly in the same way as that of the standard $\Lambda CDM$ model. In addition to dark matter and dark energy we also get another component of dark radiation. Therefore, detection of excess radiation in the universe can distinguish our model from the standard cosmology. Presence of dark radiation have its effects on primordial helium fraction and effective neutrino number. We show that our model is fully consistent with the present observational data. Future studies on the model may be able to distinguish our model from the standard cosmology.

\section*{Acknowledgment}
I wish to thank Krishnamohan Parattu, Prof. J.V.Narlikar and Prof. Tarun Souradeep for going through the draft and several interesting discussions.

\bibliographystyle{JHEP}
\bibliography{reference}

\end{document}